\documentclass[%
 reprint,
showpacs,preprintnumbers,
 amsmath,amssymb,
 aps,
prbe,
]{revtex4-1}

\usepackage{graphicx}
\usepackage{dcolumn}
\usepackage{bm}

\begin{document}

\title{A New Approach to Real Space Renormalization Group Treatment of Ising Model for Square and Simple Cubic Lattice}

\author{Tuncer Kaya}%
\email{tkaya@yildiz.edu.tr}%
\affiliation{Physics Department, Y{\i}ld{\i}z Technical University, 34220 Davutpa\c sa-Istanbul/Turkey\\
              }

\begin{abstract}
Real Space Renormalization Group (RSRG) treatment of Ising model
for square and simple cubic lattice is investigated and critical
coupling strengths of these lattices are obtained. The
mathematical complications, which appear inevitable in the
decimated partition function due to Block-spin transformation, is
treated with a relevant approximation. The approximation is based
on the approximate equivalence of $\ln(1+f(K,\{\sigma_{n.n}\}))
\simeq f(K,\{\sigma_{n.n}\})$ for small $f(K,\{\sigma_{n.n}\})$,
here $K$ is the nearest neighbor coupling strength and
$\{\sigma_{n.n}\}$ is the nearest neighbor spins degrees of
freedom around a central spin. The values of the critical coupling
strengths are obtained as $0.4830$ for square lattice and $0.2225$
for simple cubic (SC) lattice. The corresponding critical
exponents values $\alpha$ and $\nu$ are also calculated within
very acceptable agreement with those values obtained from
numerical works.

\end{abstract}

\maketitle

\section{\label{sec:level1}Introduction}

The renormalization group (RG) method accounting for large scale
fluctuations was first propounded by Kadanoff  \cite{kadanoff0}
and subsequently developed by Wilson
\cite{wilson,wilson1,wilson2,wilson3} and others into a powerful
calculation tool in the investigation of second order phase
transitions. Wilson's RG method is very general and has wide
applicability extending well beyond the field of phase
transitions. The application of RG method to a lattice system,
especially to Ising model, however, needs to have a special care.
The method developed for this purpose is known as real space
renormalization group (RSRG) which can be viewed as an
modification and extension of RG method with Kadanoff's
phenomenological ideas \cite{kadanoff,kadanoff1,ma,amit}.
Although, the references \cite{efrati,Nightingale,Plascak} include
a very nice review of historical development of RGRG methods, we
still would like to mention some of the important developments of
the method in this paper.

The fundamental idea of the RSRG approach to critical phenomena is
to calculate an equivalent form of the original partition function
by thinning out its degrees of freedom \cite{binney}. To this end,
a procedure, known as Block-spin transformation, is set up where
in each step a certain fraction of degrees of freedom is summed
up. Upon such a transformation a new decimated lattice system
appears, similar to the original one, but with fewer degrees of
freedom and with a different, renormalized, interaction constant.
The mapping of the original interaction constant, say, $K$, onto a
renormalize one $K^{'}$ constitutes the RSRG method.

As it is a half century-old method in the treatment of Ising model
and a large number of approaches have been developed and
discussed, it is difficult to give an overall view of all the
different approaches in a research paper. We would like, however,
to mention a few important developments. The finite lattice, the
cluster calculation, the cumulant expansion
\cite{niemeijer,bell,droz} and cluster variational methods
\cite{hecht} are the most commonly used approaches. The main
purpose of all these approaches or approximations are to estimate
better values for the critical parameters. For numerical studies,
the Monte Carlo renormalization group approach is a systematic
procedure for computing critical properties of lattice spin models
\cite{swedsen,swedsen1,binder,ma1,ron}.

The application of this seemingly simple method to Ising systems
creates some complications in obtaining the renormalized coupling
parameter and no further progress can be made without introducing
some sort of an approximation or truncation of some terms
appearing in the decimated partition function. It is not unusual
to solve a physics problem with an approximation. What is
interesting is that, the error involved in such approximation is
generally unknown. That is, a fundamental understanding of the
nature of these approximations has not yet been obtained.

To make this last point clear, let us consider 1D Ising model
first. In the 1D case, the decimated partition function involves
the similar type of sum that appeared in the original partition
function. On the other hand, the decimated partition function does
not involve the same type of sum that appeared in the original
partition function. Hence, it is necessary to make some
approximation to obtain renormalized coupling relation, and this
requires considerable physical intuition.

The main purpose and motivation of this paper is to treat some
Ising systems with a more followable and tractable approximate
manner rather than the approach considered in \cite{Kad}. We do not
produce new concepts or conjectures in the present paper. We
simply apply the known RSRG procedure applied to the 2D square
lattice as in the reference [23] with different approximate
considerations. We also apply the same procedure to the 3D simple
cubic Ising system. Our treatment is based on the simple fact:
when Block-spin transformation is carried out in two dimensions
and more, the decimated system contain some higher-order
interaction terms which are not present in the original system.
The renormalized coupling parameter is, then, obtained either by
omitting the higher order interactions terms totally or applying
further Block-spin transformation with the hope that some of the
parameters appearing in the decimated system can be easily
omitted. But, it is not easy to assess the quality of both of
these procedures. Therefore, a more relevant approximation in the
decimated system may produce more accurate results. As we will see
in the following chapter, the way approximations used used in this
paper not only produce a viable approximation but also produce
better estimation of the values of critical coupling strengths and
critical exponents for the Ising systems considered.

As pointed out in this section, a significant amount of different
approximations and methods have been developed and worked out in
the treatment of Ising systems from the RSRG perspective. One may
ask, then, ask the question "what is the relevance and
significance of introducing a new one". First, the approximation
used in this present paper is both simple and followable. Second,
it not only provides a viable approximate manner but also produces
better predictions for the values of the critical exponent and
critical coupling constant without introducing new concepts and
methods. Third, it is always nice to revisit an old problem from
scratch and from different aspects.

The structure of this paper is as follows. In the next section, we
are going to calculate the critical coupling strengths and
critical exponents of 2D square and 3D simple cubic Ising models.
To this end, the renormalized coupling strengths of these lattices
are going to be obtained by the usual RSRG method in a new
approximate manner. We also discuss the relevance of the treatment
used in the present paper. We also give some comparisons with the
estimations of other works.

\section{\label{sec:level1}RSRG treatment of Ising model in 2D square lattice}

Before starting our RSRG treatment, it might be better to point
out  some of the important developments in the treatment of 2D
square lattice. In two dimensions, exact critical temperature for
square lattice was estimated by Kramer and Wannier \cite{Kramer}.
Shortly afterwards, Onsager \cite{Onsager,Pott,Domb,Baxter}
determined the free energy exactly by using transfer matrix method
with periodic boundary conditions in the absence of external
magnetic field and thereby established the nature of specific-heat
singularity. The singularity in heat-capacity was interpreted as
the indication of a phase transition at a finite temperature.
Shortly afterwards, Yang \cite{yang} obtained the partition
function of 2D square lattice in the presence of external field.
The calculation carried by both Onsager and Yang is a very
complicated and cumbersome application of the transfer matrix
method. However, as we are going to see, the RSRG method  may be
the easiest viable mathematical approach in the treatment of the
2D square lattice.

The Hamiltonian of 2D square lattice Ising  model for nearest
neighbor interactions in the absence of external magnetic field is
written as follows,
\begin{eqnarray}
H(\{\sigma_{i,j})=-J\sum_{n.n}\sigma_{i,j}(\sigma_{i+1,j}+\sigma_{i,j+1}),
\end{eqnarray}
where $J$ denotes the nearest neighbor interaction constant and
corresponding partition is expressed as,
\begin{eqnarray}
Z=\small{\sum_{\{\sigma_{i,j}\}}\Huge{e}^{\tiny{K\sum\sigma_{i,j}}(\sigma_{i+1,j}+\sigma_{i,j+1})}},
\end{eqnarray}
where, $K=J/(k_{B}T)$, is the modified coupling constant (or it is
shortly called as coupling constant), $k_{B}$ and $T$  denote the
Boltzmann's constant and the temperature of the system
respectively.

Now, let us  define the spin configuration space
$\{\sigma_{i,j}\}$ as
$\{\sigma_{i,j}^{'},\sigma_{i,j+1},\sigma_{i,j-1},\sigma_{i+1,j},\sigma_{i-1,j}\}$
for the application of the RSRG method. Then, the partition
function turns out to be
\begin{eqnarray}
\
 \  \  \  \  \     \     \  \small{Z=\negthinspace\negthinspace\negthinspace\small{\small{
 \sum_{\negthinspace\negthinspace\{\negthinspace\sigma_{i,j},\sigma_{i,j}^{'}\}}}\Large{e}^{\small{K\sum}\sigma_{i,j}^{'}
(\sigma_{i+1,j}+ \sigma_{i,j+1}+\sigma_{i-1,j}+\sigma_{i,j-1})}}}.\
\
\end{eqnarray}
Now we can sum over all ${\{\sigma_{i,j}^{'}\}}$ spins and obtain
the following equivalent expression for the partition function,
\begin{eqnarray}
\small{Z=\small{\sum_{\{\sigma_{i,j}\}}\small{2^{N/2}}\LARGE{\LARGE{e}}^{\small{\sum}\ln
\cosh \small{K}((\sigma_{i+1,j}+
\sigma_{i,j+1}+\sigma_{i-1,j}+\sigma_{i,j-1})}}}.\   \   \  \
\end{eqnarray}
Now, we would like to find a renormalization transformation (or
Kadanoff transformation) that turns the partially summed partition
function into a form that looks just like the original unsummed
form. From  earlier works, we know that this is not quite possible
due to the four spins interaction terms appearing in the partially
summed (or decimated) partition function \cite{kada,pathria}.
Therefore, an inevitable approximation is necessary to obtain the
renormalized coupling constant. In this work, instead of making
this inevitable approximation after obtaining exact decimated
partition function, we make the approximation from the beginning
in the following manner. Before proceeding further, using a short
hand notation for the spin degrees of freedom may be appropriate.
Thus, we are going to use the following notation
$\sigma_{i+1,j}=\sigma_{1},\ \sigma_{i,j+1}=\sigma_{2}, \
\sigma_{i-1,j}=\sigma_{3}$, and $ \sigma_{i,j-1}=\sigma_{4}$.

Now let us to consider the logarithmic function inside the
decimated partition function with the new notation. It can be
expressed as
\begin{eqnarray}
T=\ln \cosh \small{K}(\sigma_{1}+
\sigma_{2}+\sigma_{3}+\sigma_{4}).
\end{eqnarray}
We use the following equivalent form,
\begin{eqnarray}
&&T=\ln [\cosh \small{K}(\sigma_{1}+\sigma_{2})\cosh
\small{K}(\sigma_{3}+\sigma_{4}) {}
\nonumber \\
& & \times(1+\tanh \small{K}(\sigma_{1}+\sigma_{2})\tanh
\small{K}(\sigma_{3}+\sigma_{4}))],
\end{eqnarray}
and the relations,
\begin{eqnarray}
&&\ln \cosh \small{K}(\sigma_{1}+\sigma_{2})=\frac{(1+\sigma_{1}
\sigma_{2})}{2}\ln\cosh(2K),{}
\nonumber \\
& & \ln \cosh \small{K}(\sigma_{3}+\sigma_{4})=\frac{(1+\sigma_{3}
\sigma_{4})}{2}\ln\cosh(2K), {}
\nonumber \\
& &
\tanh(\sigma_{1}+\sigma_{2})=\frac{\sigma_{1}+\sigma_{2}}{2}\tanh(\small{2K}),
{}
\nonumber \\
& & \tanh
K(\sigma_{3}+\sigma_{4})=\frac{\sigma_{3}+\sigma_{4}}{2}\tanh(\small{2K}),
\end{eqnarray}
where $\sigma_{i,j}$ takes only $\pm 1$ values. Then the
logarithmic function in Eq. (5) can be easily written as
\begin{eqnarray}
T=&&\ln\cosh(2K)+\frac{1}{2}(\sigma_{1}\sigma_{2}+\sigma_{3}\sigma_{4})\ln
\cosh(2K) {}
\nonumber \\
& &
+\ln[1+\frac{1}{4}(\sigma_{1}+\sigma_{2})(\sigma_{3}+\sigma_{4})\tanh^{2}(2K)].
\end{eqnarray}
Notice that, if all the terms in the Taylor expansion of the last
term are kept, one can obtain the same exact decimated partition
function as calculated by Kadanoff \cite{Kad}. But, the exact
decimated partition function includes four spins interaction which
leads eventually to make some inevitable approximations for
obtaining the renormalized coupling transformation. In the
Kadanoff's work, the four spin term is simply ignored without
giving any physical justification. Finally the renormalization
recursion relation, $K^{'}=(3/8)\ln \cosh(4K)$ is obtained. We,
rather, think that it would be better if renormalization recursion
relation is obtained in the following approximate manner.

Since $\tanh^{2}(2K)$ assumes small values for small values of
$K$, the first term in the expansion of the logarithmic function
might be a good approximation for small values of $\tanh^{2}(2K)$.
Apparently, the validity of our approximation depends on the
values of $\tanh^{2}(2K)$. A simple investigation shows us that if
$K<0.5$, then the approximation can be considered as a good one.
We also want to stress that, the approximation becomes better if
$K$ takes smaller values. The renormalized partition function in
Eq. (4) can be written as
\begin{eqnarray}
\small{Z=\sum_{\{\sigma_{i,j}\}}\small{f_{1}}\LARGE{\LARGE{e}}^{I_{1}+I_{2}}}
\end{eqnarray}
where, $f_{1}$, $I_{1}$ and $I_{2}$ can be expressed as
follows,
\begin{eqnarray*}
&&f_{1}=[2\cosh(2K)]^{N/2}, {}\nonumber \\&&
\small{I_{1}=\frac{1}{2}\ln \cosh(2K) \sum(\sigma_{i+1,j}
\sigma_{i,j+1}+\sigma_{i-1,j}\sigma_{i,j-1})}, {}\nonumber
\\&&I_{2}=\frac{1}{4}\tanh^2(2K)\sum(\sigma_{i+1,j}+\sigma_{i,j+1})(\sigma_{i-1,j}+\sigma_{i,j-1}).
\end{eqnarray*}
As pointed out, the last equation is just valid for small values
of $K$, and therefore it does not include four spin interaction
term. It is easy to define the following relation between the
decimated partition function and its equivalent form as,
\begin{eqnarray}
\small{\sum_{\{\sigma_{i,j}\}}\LARGE{\LARGE{e}}^{I_{1}+I_{2}}}\cong\sum_{\{\sigma_{i,j}\}}
\LARGE{e}^{K^{'}\sum\sigma_{i,j+1}^{'}(\sigma_{i-1,j}+\sigma_{i+1,j})},
\end{eqnarray}
where all $\sigma$ take $\pm 1$ values. If the the sum of the
nearest neighbor spins coupling in $I_{1}$ and the sum of the
next-nearest neighbor spins coupling in $I_{2}$ are assumed to be
approximately equal for all configurations, this equation leads to
the following renormalization transformation relation
\begin{eqnarray}
2K^{'}=\ln \cosh(2K)+\tanh^2(2K).
\end{eqnarray}
This equation has a nontrivial fixed point. That is, a finite
$K_{c}$ exists for which
\begin{eqnarray}
2K_{c}=\ln \cosh(2K_{c})+\tanh^2(2K_{c}).
\end{eqnarray}
Eq. (11) and Eq. (12) are RSRG equations that can be solved to
predict the thermodynamic properties of the 2D square lattice
Ising model. Indeed, the values of critical coupling strength can
be estimated quite readily as $K_{c}=0.483$.

The  critical exponent $\alpha$ which describes how the specific
heat diverges with temperature as we approach the critical point
$K_{c}$ from above can be calculated from the Taylor expansion of
the free energy which is equal to $N^{-1} \ln Q$ near the critical
point. Knowing that the specific heat diverges as $C\propto\mid
T-T_{c} \mid^{-\alpha}$, after some algebra, the following
relation can be obtained quite easily
\begin{eqnarray}
\alpha = 2-\frac{\ln2}{\ln (\frac{dK^{'}}{dK})\mid_{K=K_{c}}}.
\end{eqnarray}
Taking the derivative from the Eq. (11) leads to
\begin{eqnarray}
\frac{dK^{'}}{dK}= \tanh(2K)+2\frac{\tanh(2K)}{\cosh^{2}(2K)}.
\end{eqnarray}
Calculating the relation for the values of $K_{c}=0.483$, we get
$1.407$. Substituting this value into into Eq. (13) gives
$\alpha=-0.028$, which is quite acceptable if the exact value
$\alpha =0$ is considered.

The exponent $\nu$ which describes the divergence of the
correlation length of the system can also be calculated easily if
the relation valid in the neighborhood of the critical point is known. It is
indeed equal to $\xi = (K-K_{c})^{-\nu}$. After linearizing
$K^{'}$ around critical point, the relation
\begin{eqnarray}
\nu = \frac{\ln 2}{2\ln (\frac{dK^{'}}{dK})\mid_{K=K_{c}}},
\end{eqnarray}
can be obtained easily. After some algebra, the values of $\nu$
obtained as $1.014$. This estimation for $\nu$ is also quite
satisfactory, since it is not very different from the exact value
$\nu=1$.

In this section, we have used the simplest and most common scheme
in the treatment of the 2D square lattice Ising model and have
obtained the critical coupling strength and two of the critical
exponents. The estimated value of the critical coupling strength
differs from the exactly calculated value by just $10$ percent.
The estimated values of critical exponents obtained in this
section differ from the exact values by at most $2$ percent.
Improving the qualities of the estimated values of the critical
quantities might be possible by keeping more terms in the
expansion of the logarithmic function. This final indication
implies that expanding the logarithmic function until it produces
four spin interactions might be the best criterion for the
application of our approximation scheme. Of course, this requires
that the critical coupling strength must assume small values as
pointed out earlier. Luckily, we already know from other studies,
even from mean fields types theories, that the critical coupling
strength values for most of the Ising systems are not very large.
Therefore, one may expect that the application of our
approximation scheme to the 3D Ising system might turn out to be a
better approximation compared to other work done in this field.
That is what we are going to test in the following section. We
will apply our approximation scheme to the 3D simple cubic lattice
Ising model.

\section{\label{sec:level1}RSRG treatment of Ising model in 3D simple cubic lattice}
In the wake of Onsager's solution to the 2D square lattice Ising
model in zero field, several attempts were made to go beyond
Onsager, for example by solving the 3D problem in zero field. But,
none of these attempts were successful. We are now sure that there
are no exact solutions for 3D Ising systems. This makes to have a
tractable and also accurate approximation scheme for this system
very important.

The Hamiltonian of the 3D simple cubic Lattice Ising model in zero
field is written as
\begin{eqnarray}
H(\{\sigma\})=-J\sum_{n.n}\sigma_{0}(\sigma_{1}+\sigma_{2}+\sigma_{3}),
\end{eqnarray}
where $J$ denotes the nearest neighbor interaction constant and
corresponding partition is expressed as,
\begin{eqnarray}
Z=\small{\sum_{\{\sigma
\}}\Huge{e}^{\tiny{K\sum\sigma_{0}}(\sigma_{1}+\sigma_{2}+\sigma_{3})}}.
\end{eqnarray}
We used the sort hand notation for the spins as
$\sigma_{0}=\sigma_{i,j,k}$, $\sigma_{1}=\sigma_{i+1,j,k}$,
$\sigma_{2}=\sigma_{i,j+1,k}$, and $\sigma_{3}=\sigma_{i,j,k+1}$,
here $i, j$, and $k$ are integer numbers in cartesian coordinates.
Now, by the block-spin transformation the system can be decimated
as with the same manner used in previous section. Writing the
partition function as
\begin{eqnarray}
Z=\small{\sum_{\{\sigma^{'},\sigma
\}}\Huge{e}^{\tiny{K\sum\sigma_{0}^{'}}(\sigma_{1}+\sigma_{2}+\sigma_{3}+\sigma_{4}+\sigma_{5}+\sigma_{6})}},
\end{eqnarray}
and summing over the spin configuration $\{\sigma^{'}\}$ leads to
\begin{eqnarray}
Z=\small{\sum_{\{\sigma \}}\Huge{e}^{\tiny{\sum\ln[2\cosh
K}(\sigma_{1}+\sigma_{2}+\sigma_{3}+\sigma_{4}+\sigma_{5}+\sigma_{6})]}},
\end{eqnarray}
where $\sigma_{4}=\sigma_{i-1,j,k}$,
$\sigma_{5}=\sigma_{i,j-1,k}$, and $\sigma_{6}=\sigma_{i,j,k-1}$.
We now consider to rearrange the logarithmic function to have a
proper form for the investigation of the renormalize coupling
strength. $M=\ln [\cosh
\small{K}(\sigma_{1}+\sigma_{2}+\sigma_{3}+\sigma_{4}+\sigma_{5}+\sigma_{6})]$
can be expressed readily in the following form
\begin{eqnarray}
&&M=\ln [\cosh
\small{K}(\sigma_{1}+\sigma_{2}+\sigma_{3}+\sigma_{4})]+\ln[\cosh
\small{K}(\sigma_{5}+\sigma_{6})] {}\nonumber \\&&+\ln[1+\tanh
\small{K}(\sigma_{1}+\sigma_{2}+\sigma_{3}+\sigma_{4})\tanh \small
{K}(\sigma_{5}+\sigma_{6})].
\end{eqnarray}
The first term on the right hand side of the last equation can be
expressed as shown in  Eq. (8). Rearranging the term
\,\,$L=\tanh\small{K}(\sigma_{1}+\sigma_{2}+\sigma_{3}+\sigma_{4})$
as
\begin{eqnarray}
L=\frac{\tanh \small {K}(\sigma_{1}+\sigma_{2})+\tanh \small
{K}(\sigma_{3}+\sigma_{4})}{1+\tanh \small
{K}(\sigma_{1}+\sigma_{2})\tanh \small {K}(\sigma_{3}+\sigma_{4})}
\end{eqnarray}
and substituting it into last term of Eq. (20), the last term of
Eq. (20) can be written as
\begin{eqnarray*}
\small{\ln[1+C]-\ln[1+\tanh K(\sigma_{1}+\sigma_{2})\tanh
K(\sigma_{3}+\sigma_{4})]}
\end{eqnarray*}
where the function $C$ equals to
\begin{eqnarray*}
&&\tanh K(\sigma_{1}+\sigma_{2})\tanh
K(\sigma_{3}+\sigma_{4})+{}\nonumber \\&& [\tanh
K(\sigma_{1}+\sigma_{2})+\tanh K(\sigma_{3}+\sigma_{4})]\tanh
K(\sigma_{5}+\sigma_{6}).
\end{eqnarray*}
Now, substituting all these relations into the Eq. (20), it leads
to
\begin{eqnarray*}
&&M=\negthinspace\frac{1}{2}\negthinspace[3\negthinspace+\negthinspace(\sigma_{1}\sigma_{2}\negthinspace+\negthinspace\sigma_{3}
\sigma_{4}\negthinspace+\negthinspace\sigma_{5}\sigma_{6})]\ln\cosh2K+{}\nonumber
\\&& \small{ \small {\ln\{1\negthinspace\negthinspace+\negthinspace\negthinspace\frac{1}{4}[(\sigma_{1}\negthinspace
\negthinspace+\negthinspace\negthinspace\sigma_{2})(\sigma_{3}\negthinspace\negthinspace+\negthinspace
\negthinspace\sigma_{4})\negthinspace\negthinspace+
(\negthinspace\sigma_{5}\negthinspace\negthinspace+\negthinspace\negthinspace\sigma_{6})(\negthinspace\sigma_{1}\negthinspace\negthinspace
+\negthinspace\negthinspace
\sigma_{2}\negthinspace\negthinspace+\negthinspace\negthinspace\sigma_{3}\negthinspace\negthinspace\negthinspace
+\negthinspace\negthinspace\sigma_{4})]\tanh^{2}\negthinspace
2K\}}}.
\end{eqnarray*}
Making the same discussion as we did for 2D square lattice, we can
keep only the first term in the Taylor expansion of the
logarithmic function to satisfy the validity of criterion
developed in the previous section. In this case, $M$ can be
written approximately as
\begin{eqnarray*}
&&M\cong\negthinspace\frac{1}{2}[3\negthinspace+\negthinspace(\sigma_{1}\sigma_{2}\negthinspace+\negthinspace\sigma_{3}
\sigma_{4}\negthinspace+\negthinspace\sigma_{5}\sigma_{6})]\ln\cosh2K+{}\nonumber
\\&& \small{ \small {\frac{1}{4}[(\sigma_{1}\negthinspace
\negthinspace+\negthinspace\negthinspace\sigma_{2})(\sigma_{3}\negthinspace\negthinspace+\negthinspace
\negthinspace\sigma_{4})\negthinspace\negthinspace+
(\negthinspace\sigma_{5}\negthinspace\negthinspace+\negthinspace\negthinspace\sigma_{6})(\negthinspace\sigma_{1}\negthinspace\negthinspace
+\negthinspace\negthinspace
\sigma_{2}\negthinspace\negthinspace+\negthinspace\negthinspace\sigma_{3}\negthinspace\negthinspace\negthinspace
+\negthinspace\negthinspace\sigma_{4})]\tanh^{2}\negthinspace
2K\}}}.
\end{eqnarray*}
By inserting the approximate form of $M$ into Eq. (19), the
partition function can be written approximately as
\begin{eqnarray}
Q \cong \small{\sum_{\{\sigma \}}
2^{N/2}(\cosh2K)^{3N/4}\Huge{e}^{{\tiny{\sum G}}}}
\end{eqnarray}
for small values of $K$, where the function $G$ is equal to
\begin{eqnarray*}
&&G=\frac{1}{2}[\sigma_{1}\sigma_{2}\negthinspace+\negthinspace\sigma_{3}
\sigma_{4}\negthinspace+\sigma_{5}\sigma_{6}]\ln\cosh2K+{}\nonumber
\\&& \small{ \small {\frac{1}{4}[(\sigma_{1}\negthinspace
\negthinspace+\negthinspace\negthinspace\sigma_{2})(\sigma_{3}\negthinspace\negthinspace+\negthinspace
\negthinspace\sigma_{4})\negthinspace\negthinspace+
(\negthinspace\sigma_{5}\negthinspace\negthinspace+\negthinspace\negthinspace\sigma_{6})(\negthinspace\sigma_{1}\negthinspace\negthinspace
+\negthinspace\negthinspace
\sigma_{2}\negthinspace\negthinspace+\negthinspace\negthinspace\sigma_{3}\negthinspace\negthinspace\negthinspace
+\negthinspace\negthinspace\sigma_{4})]\tanh^{2}\negthinspace
2K\}}}.
\end{eqnarray*}
Now, the renormalized coupling strength can be obtained by the
following equation,
\begin{eqnarray}
\small{\sum_{\{\sigma \}}\Huge{e}^{{{ }\tiny{\sum
G}}}}\cong\sum_{\{\sigma \}}\Huge{e}^{K^{'}\sum
\sigma_{6}(\sigma_{1}+\sigma_{5}+\sigma_{3})},
\end{eqnarray}
and this last equation produces the following renormalized coupling
strength,
\begin{eqnarray}
3K^{'}=\frac{3}{2} \ln \cosh 2K+3\tanh^{2}(2K).
\end{eqnarray}
This equation produces the following relation to obtain the
critical coupling strength for the 3D simple cubic lattice Ising
model.
\begin{eqnarray}
K_{c}=\frac{1}{2} \ln \cosh 2K_{c}+\tanh^{2}(2K_{c}).
\end{eqnarray}
From this equation the value of the critical coupling strength can
be easily obtained as $K_{c}=0.2225$. The value of the critical
coupling strength is calculated as $K_{c}=0.2216$ from the Monte
Carlo simulations and series expansion types of numerical
calculations \cite{Landau1,Landau,Bloete}. Both of these methods
are considered as methods whose results one can trust. Our
estimated critical coupling value differs from their numerical
estimations by just less then $1$ percent. It might be the best
estimation obtained from any analytical treatments so far. The
agreement of the estimated critical coupling strength and the
estimated value from the computer simulations goes well beyond
even our expectations. In addition, it might indicate the
relevance of our approximation scheme. If the criterion of our
approximation scheme developed in this paper is recalled, one can
see that the nice agrement is the result of validity of the
criterion, which claims that Eq. (23) becomes a really good
approximation if the $K$ values are small enough. In other words,
the less the $K_{c}$ values are, the best approximation the Eq.
(23) turns out to be.

As in the case of 2D square lattice we also want to calculate the
critical exponents $\alpha$ and $\nu$. If the necessary derivative
$\frac{dK^{'}}{dK}|_{K=K_{c}}$ is calculated then $\nu$ can be
obtained from Eq. (15) as $\nu=0.591$. And using the relation
$\alpha=2-3\nu$, the value of $\alpha$ is obtained as $0.226$.
These numerical estimations of the values of $\nu$ and $\alpha$
are approximately $0.63$ and $0.12$ respectively
\cite{Butera,Pelissetto}. Our estimation for those values differ
almost $10$ percent and $40$ percent from the numerically obtained
values. Although, the deviation of estimated values from the exact
values are plausible, they are not in the range of our
expectation.

In conclusion, it might be worthwhile to stress the general
feature of our simple approximation scheme developed in this
present work. The motivation of writing this paper is due to the
lack of a tractable approximation scheme in the treatment of Ising
system in the RSRG perspective. Of course, it is inevitably
necessary to use some sort of approximation scheme in the
treatment of RSRG. The main idea in making our approximation
scheme tractable is to use the approximation scheme valid for
small values of the coupling parameter $K$. The function,
$\ln(1+f(K,\{\sigma_{n.n}\}))$, appearing in the Block-spin
transformed partition function is approximated by the first term
of the Taylor expansion of the function as $f(K,\{\sigma_{n.n}\})$
for small $f(K,\{\sigma_{n.n}\})$. This is the only criterion of
our approximation. This means, if the value of $\tanh K$ is small,
our approximation works better. In deed, the values of the
critical coupling strengths are obtained as $0.4830$ for square
lattice and $0.2225$ for simple cubic (SC) lattice.

We believe that our approximation works even better for body
centered and face centered cubic lattice Ising models if their non
isomorphic Block-spin transformation nature are handled properly.
Work on these cases are in progress. Presently, we think that it
is not mature enough to include in a research paper. One can
consider, this last remark as an important open problem to work on
with the approximation scheme developed in this paper.

\end{document}